\documentclass[aps,prev,twocolumn,preprintnumbers,floatfix,nofootinbib]{revtex4-1}
\pdfoutput=1
\usepackage{graphicx}
\usepackage{bm}
\usepackage{times}
\usepackage{slashed}
\usepackage{color}
\usepackage{slashed}
\usepackage{amsmath}
\usepackage{amsthm}
\usepackage{subfigure}

\newcommand{\gmu}{$g-2_{\mu}$}
\newcommand{\be}{\begin{equation}}
\newcommand{\ee}{\end{equation}}
\newcommand{\bea}{\begin{eqnarray}}
\newcommand{\eea}{\end{eqnarray}}

\begin{document}

\title{On the Connection of \gmu , Electroweak, Dark Matter and Collider Constraints on 331 Models}

\author{Chris Kelso$^{a,b}$}
\email{ckelso@unf.edu}

\author{H.N. Long$^{c}$}
\email{hnlong@iop.vast.ac.vn}

\author{R. Martinez$^{d}$}
\email{remartinezm@unal.edu.co}

\author{Farinaldo S. Queiroz$^{e,f}$}
\email{fdasilva@ucsc.edu}

\affiliation{$^a$Department of Physics and Astronomy - University of Utah - Salt Lake City, UT 84112\\ 
$^b$Physics Department - University of North Florida - Jacksonville, FL 32224\\
$^c$Institute of Physics VAST  -10 Dao Tan - Ba Dinh, Hanoi, Vietnam\\
$^d$Departamento de Fisica - Universidad Nacional de Colombia, Bogota D.C\\
$^e$Department of Physics - University of California Santa Cruz - Santa Cruz, CA 95064\\
$^f$Santa Cruz Institute for Particle Physics, Santa Cruz, CA 95064\\}

\begin{abstract}
In this work we compute all contributions to the muon magnetic moment stemming from several 3-3-1 models namely,
 minimal 331, 331 with right handed neutrinos, 331 with heavy neutral leptons, 331 with charged exotic leptons,
 331 economical and 331 with two higgs triplets. Further, we exploit the complementarity among current electroweak,
 dark matter and collider constraints to outline the relevant parameter space of the models capable of
 explaining the anomaly. Lastly, assuming that the experimental anomaly has been otherwise resolved, we derive robust $1\sigma$ bounds using the current and projected measurements.
\end{abstract}


\maketitle

\section{Introduction}
\label{intro}

The muon magnetic moment (\gmu) is one of the most precisely measured quantities in particle physics. Somewhat
recently in Brookhaven, \gmu\, has been measured with great a precision reaching the level of 0.54 ppm.
Since the first results were reported, a long standing discrepancy between theory and experiment
of about $3.6\sigma$ has been observed, providing a hint that new physics may be around the corner.
This deviation triggered a multitude of speculations about the possible origin of this mild excess
(for recent reviews see Refs.\cite{muonreview_1,muonreview_2}). However, there are large theoretical uncertainties
 that blur the significance of this discrepancy. These uncertainties are dominated by the hadronic vacuum polarization and the
 hadronic contribution to the light-by-light scattering. Significant effort has been put forth to try to reduce these uncertainties \cite{g2improvements1,g2improvements2,g2improvements3}.  The current deviation is
 $\Delta a_{\mu}=295 \pm 81 \times 10^{-11}$. Out of this $\pm 81 \times 10^{-11}$ error, $\pm 51
 \times 10^{-11}$ is theoretical, which is dominated by uncertainty in the lowest-order hadronic contribution
  ($\pm 39  \times 10^{-11}$) and in the hadronic light-by-light contribution ($\pm 26  \times 10^{-11}$)
  \cite{fermilabproposal}.

In the near future important improvements in both the theoretical and experimental situations are expected.
Combining the expected progress from the theoretical side, along with the projected experimental sensitivity
for the g-2 experiment at Fermilab, the precision will likely reach $\Delta a_{\mu}=295 \pm 34 \times 10^{-11}$,
possibly increasing the magnitude of the signal up to $5\sigma$ \cite{fermilabproposal}. Hence, it is worthwhile
to explore the complementarity among \gmu, electroweak, dark matter and collider constraints in particle physics models.

In this work, we will focus our effort on electroweak extensions of the standard model known as 331 models.
In these models the $SU(2)_L$ gauge group is extended to $SU(3)_L$. The motivations for considering such class of models,
among others \cite{othermotiv_1,othermotiv_2,othermotiv_3}, relies on the following:\\
(i) They explain the number of generations: the
331 gauge symmetry in combination with QCD asymptotic freedom lead to
the generation number to be three;\\
(ii) They have plausible dark matter candidates \cite{DM331_1,DM331_2,DM331_4,DM331_7,DM331_8,DM331_9,DM331_10,DM331_11,DM331_12};\\
(iii) They can accommodate the dark radiation component observed by Planck through non-thermal DM production \cite{DM331_3}\\
(iv) They are generally consistent with current electroweak and collider data as we discuss further;\\
(v) The Peccei-Quinn symmetry, necessary to solve the strong-CP
problem, follows naturally from the particle content in these models
\cite{cp331}.

Our goal is to assess which 331 models are consistent with the current electroweak, collider and dark matter
limits while being able to explain the \gmu\, anomaly and derive $1\sigma$ bounds on the particle spectrum.
Previous studies have been performed in the past discussing the \gmu\, in 331 models \cite{muon331_1,muon331_2}.
Those studies were limited to one particular model, such as the minimal 331 model and 331 model with right handed neutrinos
 without taking into account important electroweak, collider and dark matter constraints. In this work we will
 extend those studies by investigating the \gmu\, in six 331 models namely: the minimal 331 model \cite{pleitezref},
 331 model with right handed neutrinos (331 r.h.n for short) \cite{331rhn}, 331 model with heavy neutral leptons
 (331LHN) \cite{DM331_1}, 331 Economical \cite{331economic}, 331 Minimal with two higgs triplets (RM331 for short),
  and the 331 with charged exotic leptons \cite{331exotic_1,331exotic_2}, properly accounting for these constraints.
  Additionally, we derive $1\sigma$ limits based on the current and projected sensitivity for \gmu assuming the anomaly has been otherwise resolved.

In summary our main findings are:

\begin{itemize}
\item {\it 331 Minimal:}
This model cannot explain \gmu anomaly. We find a robust limit on the scale of symmetry breaking ($v_{\chi}$) of $4$~TeV,
which can be translated into $M_{Z^{\prime}} > 2.4$~TeV. As far as we know this is the strongest bound
on the $Z^{\prime}$ mass in the literature. Moreover, we show that the upcoming g-2 experiment at
Fermilab might be able to fiercely exclude this model.

\item {\it 331 r.h.n:}
It cannot explain \gmu\, excess because it requires a rather small scale of symmetry breaking already ruled out by current collider, dark matter experiments and electroweak precision data.

\item {\it 331 LHN:}

For heavy neutrino masses of $M_N=1$~GeV, $v_{\chi} < 1$~TeV is needed to address \gmu. Nevertheless, current bounds prohibit this possibility. Hence, the 331LHN is excluded as a potential framework. Because the overall contribution is quite small, a projected $1\sigma$ limit of $v_{\chi} \gtrsim 1.5$~TeV is somewhat irrelevant compared to the current
direct dark matter detection ones \cite{DM331_4,DM331_7}. The regime in which the heavy neutrino masses are either larger or smaller do not change our conclusions.

\item {\it 331 Economical:}
Due to the large cancelation between the $W^{\prime}$ and
$Z^{\prime}$ corrections the total contribution is small and
requires $v_{\chi} < 1$~TeV to accommodate the \gmu\  excess.
Such a low scale of symmetry breaking is prohibited by current data, however. In
conclusion, the Economical 331 model cannot reproduce the \gmu\ reported, no meaningful current limit can be derived, but a projected one of $1.4$~TeV is found.

\item {\it RM331:}
We observe that a scale of symmetry breaking of $\sim 2$~TeV could explain the \gmu\, excess,
while being consistent with existing limits. Furthermore, a current limit of 4~TeV and projected limit of $6$~TeV can be placed on the scale of symmetry
 breaking. Since this model, similar to the 331 minimal model, is valid up to only 5 TeV, the RM331 may be ruled out in the near future.

\item {\it 331 Exotic Leptons:}

Regardless how massive the exotic leptons are, the total contribution to \gmu\, is negative and small. Therefore, no relevant constraint could be derived.

\end{itemize}

We have given a brief introduction to the current status of the muon anomalous magnetic moment and
summarized our main findings. We now turn our attention to the corrections to \gmu\, stemming from
the most popular 331 models. Throughout this work, our reasoning focuses on the leptonic sectors of such models as they are the most relevant for the \gmu\, anomaly.  We also properly account for the existing electroweak and collider bounds on the
other particles of the models. We provide master integrals and analytical
expressions for all contributions to \gmu\, discussed in this work in the Appendix.

\section{Minimal 331 Model}

\subsection{Content}
The leptonic content of the minimal 331 model is comprised of three lepton triplets as follows,
\be
f^{a}_L = \left(
               \nu^a,\  l^a,\ (l^c)^a
                 \right)_L^T \sim (1, 3, 0),
\label{l}
\ee
where {\it a} runs through the three family generations. Since 331 stands for an enlarged electroweak gauge
symmetry, 5 new gauge bosons are added to the SM namely, $W^{\prime \pm}$, $U^{\pm \pm}$ and $Z^{\prime}$.
Both $W^{\prime \pm}$ and $U^{\pm \pm}$ carry two units of lepton number, hence called bileptons, which
interact with the SM leptons as follows \cite{pleitezref},
\be
{\cal L}^{CC}_l \supset - \frac{g}{2\sqrt{2}}\left[
\bar{\nu}\gamma^\mu (1- \gamma_5) C\bar{l}^{T}W^{\prime -}_\mu
 -  \bar{l}\gamma^\mu \gamma_5 C \bar{l}^T
U^{--}_\mu +h.c\right],
\label{doublyminimal}
\ee with the respective masses,

\begin{eqnarray}
M_{W^{\prime}}^2 = \frac{g^2}{4}\left( v_{\eta}^2 + v_{\chi}^2 +
v_{\sigma}^2 \right), M_U^2 =\frac{g^2}{4}\left( v_{\rho}^2 + v_{\chi}^2
+ 4 v_{\sigma}^2 \right),\nonumber\\ 
\label{massWprime}
\end{eqnarray}where $v_{\rho},v_{\eta}$,  $v_{\chi}$ and $v_{\sigma}$ are the vev's of the neutral scalars presented in the Eq.(\ref{tripletscalars}) below.

\begin{figure*}[!t]
\centering
\subfigure[\label{feymann6}]{\includegraphics[scale=0.6]{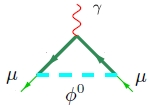}}
\subfigure[\label{feymann1}]{\includegraphics[scale=0.6]{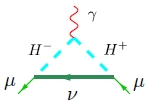}}
\subfigure[\label{feymann9}]{\includegraphics[scale=0.6]{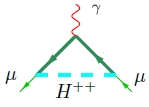}}
\subfigure[\label{feymann8}]{\includegraphics[scale=0.6]{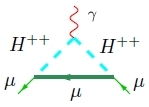}}
\subfigure[\label{feymann2}]{\includegraphics[scale=0.6]{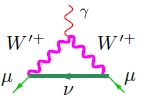}}
\subfigure[\label{feymann3}]{\includegraphics[scale=0.6]{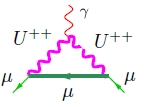}}
\subfigure[\label{feymann4}]{\includegraphics[scale=0.6]{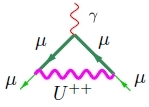}}
\subfigure[\label{feymann5}]{\includegraphics[scale=0.6]{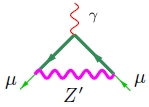}}
\subfigure[\label{feymann7}]{\includegraphics[scale=0.6]{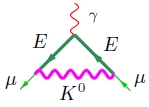}}
\subfigure[\label{feymann10}]{\includegraphics[scale=0.6]{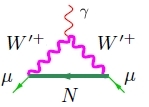}}
\caption{Feynmann diagrams arising in 331 models studied here.}
\end{figure*}

Notice that the vector current for $U^{\pm \pm}$ vanishes due to
Fermi statistics.  One can clearly see that both charged bosons
generate contributions to $g-2_{\mu}$ through Fig.\ref{feymann2} and Figs.\ref{feymann3}-\ref{feymann4}, respectively. Regarding the neutral gauge
boson $Z^{\prime}$, which mixes with the SM  $Z$, we find the
gauge interactions \cite{pleitezref}, \be {\cal L}^{NC} \supset
\bar{f}\, \gamma^{\mu} [g_{V}(f) + g_{A}(f)\gamma_5]\, f\,
Z'_{\mu}. \label{ncm} \ee with, \bea g_{V}(\mu) & &= \frac{g}{c_W}
\frac{3\sqrt{1 - 4 s_W^2}}{2\sqrt{3}},\
g_{A}(\mu) = \frac{g}{c_W} \frac{\sqrt{1 - 4 s_W^2}}{2\sqrt{3}},\nonumber\\
& &M_{Z^{\prime}}^2 = \left( \frac{g^2+3g^{\prime 2}}{3} \right) v_{\chi}^2,
\label{hsm}
\eea where $g^{\prime} = g \tan_W$. Electroweak measurements constraint this mixing angle to be quite small \cite{ZZpmixing}. Note the $Z^{\prime}$ boson also contributes to the muon anomalous magnetic moment through the diagram shown in Fig.\ref{feymann5}. As we will see later, the magnitude of the vector and
axial couplings lead to the $Z^{\prime}$ correction to $g-2_{\mu}$ that is positive in this model. The $Z^{\prime}$ contribution
to \gmu\, in general is proportional to $g_V^2 - 5g_A^2$ where $g_V$ and $g_A$ are the vector and axial
couplings. Therefore, depending on the hypercharges assigned for the leptonic triplets, which
determine $g_V$ and $g_A$, the magnitude and possibly the overall sign of the $Z^{\prime}$ contribution to \gmu\,  can change
(see Appendix for details).

As for the scalar sector, 331 models usually advocate the presence
of three scalar triplets and one sextet in order to
generate masses for all fermions. In the case of the minimal 331
model those read,

\bea \eta & = & \left(
               \eta^0,\  \eta^+_1,\ \eta_2^+
                 \right)^T,\nonumber\\
\rho & = &\left(
               \rho^+,\  \rho^0,\ \rho^{++}
                 \right)^T,\nonumber\\
\chi & = & \left(
               \chi^-,\  \chi^{--},\ \chi^0
                 \right)^T,\nonumber\\
S  & = & \left( \begin{array}{ccc}
\sigma_1^0 & h_2^- & h_1^+\\
h_2^- & H_1^{--} & \sigma_2^0\\
h_1^+ & \sigma_2^0 & H_1^{++}
\label{tripletscalars}
 \end{array}  \right),
\eea where $\eta^0, \rho^0, \chi^0$ and $\sigma_1^0$ acquire a vev $v_{\eta},v_{\rho},v_{\chi}$ and $v_{\sigma}$ respectively.
 
The important interactions for muon magnetic moment are \cite{Foot:1992rh}:

\be {\cal L} \supset G_l\,\left[ \overline{l_R}\, \nu_L \eta_1^- + \overline{l_R^c}\, \nu_L h_1^+ + \overline{l_R}\, \nu_L h_2^+ + \overline{l_R} l_L R_{\sigma_2} \right]
+ h.c 
\label{chargedscal}
\ee
 with \cite{Tonasse:1996cx},

\bea
 M_{\eta_1^+}^2 & = & \frac{f }{\sqrt{2}}v_{\chi} \left(
\frac{v_{\rho}}{v_{\eta}} + \frac{v_{\eta}}{v_{\rho}} \right),\nonumber\\ 
M_{h_1^+,h_2^+} & & \sim v_{\chi} ,\nonumber\\
 M_{\sigma_0^{\prime}} & &\sim v_{\chi},
\eea 
where $G_l = m_l \sqrt{2}/v_{\eta}$, with $v_{\eta}$ being the vev of $\eta^0$, f the trilinear coupling in the scalar potential \cite{pleitezref} and $R_{\sigma_2}$ the real component field of $\sigma_2^0$ \cite{Tonasse:1996cx}. After the spontaneous symmetry
breaking we find $v_{\eta}^2 + v_{\rho}^2 + v_{\sigma}^2= v^2$ where $v$ is the SM vev. Typically $v_{\sigma}$, which gives rise to neutrino masses, is taken to be small, whereas $v_{\eta}$ is assumed to be equal to $v_{\rho}$, but $v_{\eta}$ is free to vary obeying this restriction, and vice-versa.
This is important because for small values of $v_{\eta}$ the charged scalar contribution will not be suppressed as has been previously assumed \cite{muon331_1}. The feynmann diagrams which give rise to correction to \gmu from these scalars are depicted in Figs.\ref{feymann6}-\ref{feymann1}.

We have shown the relevant interactions to \gmu\, thus far, further we discuss the existing
constraints on the minimal 331 model.

\subsection{Existing Bounds}

Since this model does not have a dark matter candidate, the important bounds arise from electroweak and
collider data only. The muon decay $\mu \rightarrow e \nu_e \bar{\nu_{\mu}}$ implies $M_{W^{\prime}} >
230$~GeV \cite{muondecay}. Measurements of flavor changing neutral current (FCNC) in meson oscillations produce a lower bound of $M_{Z^{\prime}} \gtrsim 1-2$~TeV. The variation in the bound comes from the
texture parametrization used in the quark mixing matrices. Additionally, electroweak
bounds coming from the rare decays $B_{s,d} \rightarrow \mu^+ \mu^-$ and $B_{d} \rightarrow K^{\star} (K)
\mu^+ \mu^-$ impose $M_{Z^{\prime}} \gtrsim 1$~TeV masses \cite{jennifer1}. Lastly, CMS Collaboration has
performed $Z^{\prime}$ searches. Since no excess has been observed, a bound of $2.2$~TeV has been found on
$Z^{\prime}$ mass \cite{LHCZprimeminimal}. This limit can be translated into a limit on the scale of symmetry breaking of the model of $v_{\chi} > 3.6$~TeV. We will incorporate these bounds in our results in the next section.

\section{331 Model with Right Handed Neutrinos}

The so called 331 model with right handed neutrinos (331 r.h.n for
short) is motivated by neutrino masses.
Here the neutrino masses can be easily addressed. This model is an extension of the minimal 331 model in which the third
component in the leptonic triplet for a right handed neutrino is replaced as follows \cite{331rhn},

\be
f^{a}_L = \left(  \nu^a, l^a,  (\nu^c)^a
\right)_L^T \sim (1, 3, -1/3), l^a_R\sim (1, 1, -1).
\label{lr}
\ee

As before five new gauge bosons are added to the SM namely, $W^{\prime},
X^{0},X^{0\dagger}$ and $Z^{\prime}$. Because in this model the third component in the leptonic triplet
is a neutral particle, this model does not feature doubly charged bosons, it has instead neutral bosons
$X^{0},X^{0\dagger}$. The spontaneous symmetry breaking (SSB) induces
$|M_{W^{\prime}}^2 - M_X^2| \leq M_W^2$ ~\cite{331rhn}, and a $Z-Z^{\prime}$ mixing. Since the $Z-Z^{\prime}$ mixture is bounded to be very small we might consider Z and $Z^{\prime}$ as mass eigenstates. In this regime the
vector and axial couplings using the notation of Eq.(\ref{ncm}) are found to be,

\be
g'_{V}(\mu) = \frac{g}{4 c_W} \frac{(1 -
4 s_W^2)}{\sqrt{3-4s_W^2}},\
g'_{A}(\mu) = -\frac{g}{4 c_W \sqrt{3-4s_W^2}},
\label{hsr}
\ee with

\bea
M_{Z^{\prime}}^2 = \frac{g^2}{4(3-4s_w^2)}\left( 4 v_{\chi}^2 + \frac{v_{\rho}^2}{c_w^2}
+ \frac{v_{\eta}^2(1-2s_w^2)}{c_w^2} \right).
\label{zprimemassmin}
\eea

The $Z^{\prime}$ boson contribution to \gmu\ appears in the form of Fig.\ref{feymann5}. The  bilepton  $X^0$ does not contribute to \gmu, but similar to the Minimal 331 model the singly charged boson does, through the interaction (Fig.\ref{feymann2}),

\be
{\cal L} \supset - \frac{g}{2\sqrt{2}}\left[
\overline{\nu^c_R}\, \gamma^\mu (1- \gamma_5) \bar{l}\, W^{\prime -}_\mu  \right].
\ee

The mass term of the singly charged gauge boson is similar to the previous model according
to Eq.(\ref{massWprime}). The scalar sector in the 331 r.h.n is different though and it
is now comprised of the following three scalar triplets,

\bea
\eta = \left(
               \eta^0,\  \eta^-,\ \eta^{0 \prime}
                 \right)^T,\nonumber\\
\rho = \left(
               \rho^+,\  \rho^0,\ \rho^{\prime +}
                 \right)^T,\nonumber\\
\chi = \left(
               \chi^0,\  \chi^{-},\ \chi^{\prime 0}
                 \right)^T.
\label{tripletscalars2}
\eea

Due to a different scalar content this model has contributions to \gmu\, stemming from three scalars.
Two coming from singly charged ones, with an interaction similar to Eq.(\ref{chargedscal}) represented in Fig.\ref{feymann1}. In addition,
there is a correction coming from a neutral scalar $S_2$, which is a combination of the real component of the $\rho^0$ and $\eta^0$ fields, exhibited in Fig.\ref{feymann6} through the interaction,

\be
{\cal L} \supset G_s \bar{\mu}\, \mu S_2,
\label{neutralSca}
\ee where $G_s=m_{\mu} \sqrt{2}/(2 v_{\rho})$, and

With those results we have gathered all information needed for the \gmu. We emphasize that the key
differences between the 331 r.h.n model and the minimal 331 are the absence of the doubly charged and
the presence of additional charged and neutral scalar contributions. Before presenting our main findings
for this particular model we discuss further the existing bounds.

\subsection{Existing Bounds}

The non-observation of an excess in the dilepton search from CMS experiment has resulted in a $2.4$~TeV lower bound on
the $Z^{\prime}$ mass \cite{LHCZprimeminimal}. This limit can be translated
into a lower bound on the scale of symmetry breaking of $7.5$~TeV using Eq.(\ref{zprimemassmin}). Electroweak
data from the decays $B_{s,d} \rightarrow \mu^+ \mu^-$ and $B_{d} \rightarrow K^{\star} (K) \mu^+ \mu^-$ exclude
$Z^{\prime}$ masses up to $\sim 1-3$~TeV \cite{jennifer1}. Additionally, direct dark matter detection bounds coming
from the underground detector LUX, have been applied to the 331 model with right-handed neutrinos to exclude
a scale of symmetry breaking lower than $10$~TeV \cite{DM331_7}, implying $M_{Z^{\prime}} \gtrsim 4$~TeV. The
latter is valid under the assumption that the complex scalar $\phi$ which is $\sim \eta^{0 \prime}$ is a viable DM candidate.  With those stringent constraints in mind we show our results concerning \gmu.

\section{331 Model with Heavy Neutral Lepton}

The 331 model with heavy leptons (331LHN for short) is a compelling extension of the SM because it can obey the electroweak constraints and has two viable dark matter candidates, a complex scalar and a fermion \cite{DM331_1} in the context of the Higgs \cite{Queiroz:2014yna} and $Z^{\prime}$ portals \cite{Alves:2013tqa} respectively. Besides, it offers a possible explanation to the dark radiation favored by current data through a sub-dominant non-thermal production of dark matter \cite{DM331_3,Allahverdi:2013noa}. Here the third component in the leptonic triplet is a heavy neutral lepton as follows,
\be
f^{a}_L = \left(  \nu^a, l^a,  N^a
\right)_L^T \sim (1, 3, -1/3), l^a_R\sim (1, 1, -1).
\label{lr}
\ee

This model is nearly identical to the 331 model with right handed neutrinos as far the muon anomalous
magnetic moment is concerned. The key differences rise from the presence of the heavy leptons represented in Fig.\ref{feymann10} through the interactions,

\bea
{\cal L} \supset - \frac{g}{\sqrt{2}}\left[
\overline{N_L}\, \gamma^\mu \bar{l}\, W^{\prime -}_\mu  \right]
-G_l\, \overline{l_R}\, N_L \eta_1^-
\label{heavylep1}
\eea

Notice the presence of the heavy lepton instead of the light neutrino in the previous model. In summary,
the interactions that contribute to the muon anomalous magnetic moment in the 331 model with heavy leptons
stem from the singly charged gauge boson Eq.(\ref{heavylep1}), the $Z^{\prime}$ with the vector and axial couplings
 of Eq.(\ref{hsr}), the neutral scalar $S_2$ via Eq.(\ref{neutralSca}), singly charged $h_1$ through Eq.(\ref{heavylep1})
  and the second singly charged scalar $h_2$ via Eq.(\ref{chargedscal}). Because of the heavy lepton, the
  singly charged scalars give rise to different corrections to \gmu\, as we shall see below.


\subsection{Existing Bounds}

The existing bounds on this model are very similar to the 331 r.h.n. CMS dilepton searches resulted in
the lower bound $M_{Z^{\prime}} \gtrsim 4$~TeV \cite{LHCZprimeminimal} in the regime where $Z^{\prime}$
boson cannot decay into heavy lepton pairs. This bound demands a scale of symmetry breaking larger than
$7.5$~TeV. Electroweak bounds coming from the $B_{s,d} \rightarrow \mu^+ \mu^-$ and $B_{d} \rightarrow
K^{\star} (K) \mu^+ \mu^-$ rule out $Z^{\prime}$ masses up to $\sim 3$~TeV \cite{jennifer1}.
Because this model has two non-coexistent dark matter candidates, a complex scalar ($\eta^{0\prime}$)
and the lightest heavy lepton ($N_1$ for instance), the dark matter bounds change depending on which
particle is the lightest. For the scenario where the scalar is the lightest, direct dark matter detection
excludes a scale of symmetry breaking lower than $10$~TeV, implying $M_{Z^{\prime}} \gtrsim 4$~TeV \cite{DM331_7}.
For the regime where the fermion is the DM candidate, it has been found $v_{\chi^{\prime}} > 5$~TeV, i.e $M_{Z^{\prime}}
 \gtrsim 2$~TeV \cite{DM331_4}. 

\section{Economical 331 Model}

This model refers to the 331 extension which uses the leptonic triplet of the 331 with
handed neutrinos, but instead of having three scalar triplets it has only two namely \cite{331economic},

\bea
\phi = \left(
               \eta_1^+,\  \eta^0_2,\ \eta_3^+
                 \right)^T,\nonumber\\
\chi = \left(
               \chi^0_1,\  \chi^{-}_2,\ \chi^0_3
                 \right)^T.
\eea

Here we will adopt the notation: $v_{\eta_2^0} = v\sqrt{2}, v_{\chi_1} = u/\sqrt{2}$
and $v_{\chi_3^0} = v_{\chi}/\sqrt{2}$, where $v$ is the SM vev.

This model possesses a more simple scalar sector compared to the 331 with right handed neutrinos.
As a result of this simplicity, corrections to \gmu\, arise from neutral and charged scalars (See Eq.(15)
of \cite{331economic}) similar to Eq.(\ref{neutralSca}) and Eq.(\ref{chargedscal}) 
 but with the following masses after replacing $v_{\eta}$ by {\it v},,

\bea
M_{\eta_1^+}^2 =\frac{ \lambda_4}{2} \left( u^2 +v^2 + v_{\chi}^2 \right), \,
M_{S_2}^2 = 2 \lambda_1 v_{\chi}^2
\eea

The diagrams that contribute to \gmu rising from these scalars are exhibited in Fig.\ref{feymann6}-\ref{feymann1}. Because the 331 Economical model has the same leptonic triplet of the 331 r.h.n , i.e, the same hypercharge
 configuration, the vector and axial $Z^{\prime}$ couplings are the same, but the $Z^{\prime}$ mass term
 turns out to be different as result of the different scalar content as follows,

\be
M_{Z^{\prime}}^2 = \frac{g^2 c^2_w v_{\chi}^2}{3-4s_w^2}.
\ee

As for the singly charged boson interaction, it is identical to the 331 r.h.n model and given in
Eq.(\ref{chargedscal}). In summary, the relevant contributions to \gmu\, from this model come from the
charged and neutral gauge bosons. 

\subsection{Existing bounds}

Data from the $B_{s,d} \rightarrow \mu^+ \mu^-$ and $B_{d} \rightarrow K^{\star} (K)
\mu^+ \mu^-$ decays exclude $Z^{\prime}$ masses up to $\sim 1-3$~TeV depending on the parametrization in the
quark mixing matrices \cite{jennifer1}.  Dilepton searches performed by CMS resulted in the lower bound
$M_{Z^{\prime}} \gtrsim 4$~TeV , which implies $v_{\chi} > 7.5$~TeV \cite{LHCZprimeminimal}. 

\section{Minimal 331 Model with two Higgs Triplets}

The minimal 331 model with two Higgs triplets, RM331 for short, was mostly motivated
by minimality due to the shortened scalar sector. This model does not have a dark matter
candidate, nor does it explain the fermion masses with renormalizable Lagrangians \cite{Ferreira:2011hm}.
The scalar sector of this model is comprised of two scalar triplets only namely,

\bea
\rho = \left(
               \rho^+,\  \rho^0,\ \rho^{++}
                 \right)^T,\nonumber\\
\chi = \left(
               \chi^-,\  \chi^{--},\ \chi^0
                 \right)^T.
\eea

As a result, contributions from doubly charged (Figs.\ref{feymann9}-\ref{feymann8}) and neutral scalars (Figs.\ref{feymann6}) arise through the Lagrangian,

\be
{\cal L} \supset \frac{m_{\mu}}{v_{\chi}} \bar{l}\, l\, S_2 + \sqrt{2}\frac{m_{\mu}}{v_{\rho}}
  \bar{l^c}\, l_L\, H^{--},
\ee with,

\bea
M_{S_2}^2 = \lambda_2 v_{\chi}^2 + \frac{\lambda_3^2 v_{\rho}^2 }{4 \lambda_2}, \,
M_{H^{\pm \pm}}^2 = \frac{\lambda_4}{2} \left( v_{\chi}^2 + v_{\rho}^2 \right).
\eea

 This model possesses a similar hypercharge configuration, vector and axial couplings to the minimal 331 model.  This means that the $Z^{\prime}$-lepton
 interactions are equivalent to those described in Eq.(\ref{hsm}), but with,

\be
M_{Z^{\prime}} = \frac{g^2 c_w^2 v_{\chi}^2}{3(1-4 s_w^2)}.
\ee

The singly charged $V^{\pm}$ (Fig.\ref{feymann2}) and doubly $U^{\pm \pm}$ (Figs.\ref{feymann3}-\ref{feymann4}) charged
vector bosons contributions are precisely the same of the minimal
331 model, Eq.(\ref{doublyminimal}), changing the mass terms only
accordingly,

\bea
M_V^2 = \frac{g v_{\chi}^2}{4}, \,\,
M_U^2 = \frac{g^2}{4}\left( v_{\rho}^2 + v_{\chi}^2 \right).
\eea

Thus, in summary, the corrections to \gmu\, stemming from this model are: neutral scalar, doubly charged scalar,
doubly charged vector boson, singly charged vector boson and neutral boson. 

\subsection{Existing Limits}

Currently limits based on Drell Yann production of doubly charged scalar exclude doubly charged
scalars up to $\sim 400$~GeV \cite{doublycharged}. The muon decay $\mu \rightarrow e \nu_e
\bar{\nu_{\mu}}$ implies $M_{W^{\prime}} > 230$~GeV \cite{muondecay}. Flavor changing neutral
current processes arising from the RM331 model are sizeable and therefore stringent constraints have
been found: $v_{\chi} \gtrsim 1-2.7$~TeV, depending on the texture parametrization
used \cite{diegoFCNC}. Moreover, data from the $B_{s,d} \rightarrow
\mu^+ \mu^-$ and $B_{d} \rightarrow K^{\star} (K) \mu^+ \mu^-$ decays rule out $Z^{\prime}$ masses
up to $1-2$~TeV range \cite{jennifer1}. In addition, the CMS Collaboration has performed $Z^{\prime}$
searches, since no excess has been observed a lower bound $M_{Z^{\prime}} \gtrsim 2.2$~TeV was
derived \cite{LHCZprimeminimal}. This limit can be translated into a limit on the scale of
symmetry breaking of the model namely $v_{\chi} > 1850$~TeV. 

\section{331 Model with exotic Leptons}

A special feature of the 331 models discussed previously is the fact
that one quark generation transforms in a different representation
of $SU(3)_L$ compared to others,  in order to satisfy the
chiral anomaly cancelation condition. As a result, the
$Z^{\prime}$-quark interactions are not universal, giving rise to
flavor changing neutral current processes at tree level
\cite{diegoFCNC}. Different 331 models can be built, in
particular some are comprised of five left-handed leptonic
triplets in different representations of the $SU(3)_L$ gauge group
\cite{different331}. Within this context the $Z^{\prime}$-lepton
interactions are not universal and flavor changing processes can arise. We investigate
these models in this section. 

The leptonic sector is comprised
of the following triplets,

\bea
f_{1L} = \left(  \nu_1, l_1^-,  E_1^-
\right)_L^T \sim (1, 3, -2/3), \, l_1^c\sim (1, 1, 1)\nonumber\\
f_{2,3L} = \left( l_{2,3}^- ,\nu_{2,3},  N_{2,3}
\right)_L^T \sim (1, 3^{\star}, -1/3), \, l_{2,3}^c\sim (1, 1, 1)\nonumber\\
f_{4L} = \left( E_2^- ,N_{3},  N_{4}
\right)_L^T \sim (1, 3^{\star}, -1/3), \, E_{2}^c\sim (1, 1, 1)\nonumber\\
f_{5L} = \left( N_5 ,E_2^+,  l_{3}^+
\right)_L^T \sim (1, 3^{\star}, 3/3), \, E_{2}^c\sim (1, 1, 1)\nonumber\\
\label{lr}
\eea
Using the same notation of Ref.\cite{Cabarcas:2013jba}, the relevant interactions for the \gmu\, are,

\bea
{\cal L} \supset \frac{g^{\prime}}{2 \sqrt{3} s_w c_w} \bar{\mu}\gamma_{\mu} \left( g_V + g_A \right) \mu \, Z^{\prime}\nonumber\\
 - \frac{g}{\sqrt{2}} \left( \overline{N_{1L}}\, \gamma_{\mu} \mu_{L} + \bar{\mu}_L
 \gamma_{\mu} N_{4L} \right) K^{+}_{\mu}\nonumber\\
 -\frac{g}{\sqrt{2}} \left( \bar{\mu}_L \gamma_{\mu} E_L \right) K^0_{\mu}\nonumber\\
h_1 \bar{\mu} (1-\gamma_5) N \phi^+ + h_2 \bar{\mu} E^- \phi^0 +h_3 \bar{\mu} E_2^- \phi^0  +h.c.,
\label{exotcontri}
\eea with,

\bea
g_V = \frac{-c_{2w} + 2 s_w^2}{2}, \, g_A=\frac{c_{2w} + 2 s_w^2}{2},\nonumber\\
M_{Z^{\prime}} =\frac{2}{9} \left( 3 g^2 + g^{\prime 2} \right) v_{\chi}^2,\nonumber\\
M_{K^+}^2 =M_{K^0}^2= \frac{g^2}{4} \left( 2 v_{\chi}^2 + v^2 \right),\nonumber\\
g^{\prime} =\frac{g \tan_w}{\sqrt{1- \tan_w^2/3}},
\eea where $K^{+},K^0$ and $Z^{\prime}$ are the gauge bosons of the model, $v_{\chi}$ sets the scale
of the $SU(3)_L$ symmetry breaking, $v$ is the SM vev, and $\phi^+, \phi^0$ are the heavy charged and
neutral scalars evoked in the scalar triplets \cite{Ponce:2001jn}. We have seen that corrections rising from 
scalar particles are suppressed by the lepton masses, so we will ignore them here. Hence the main contributions
come from the gauge bosons $Z^{\prime}$ (see Fig.\ref{feymann5}), $K^+$ (see Fig.\ref{feymann2}) and $K^0$ (check Fig.\ref{feymann7}) . 

\subsection{Existing Constraints}

A study of rare decay data, $\tau \rightarrow lll, \tau \rightarrow e \gamma, \mu \rightarrow e \gamma$
and $\mu \rightarrow eee$, has placed limits on the $Z^{\prime}$ mass which range from 800~GeV-4~TeV,
depending on the value of the mixing angles in the leptonic sector. This implies a lower bound
$v_{\chi} \gtrsim 1.5 -7.1$~TeV. Constraints coming from pair production of charged exotic leptons (E) at the LHC imply $M_E > 405$~GeV \cite{Freitas:2014pua}.  This can be effectively translated into a limit on the scale of symmetry breaking of the model, since it depends on the product of Yukawa coupling and $v_{\chi}$ and the Yukawa coupling can be arbitrarily small. 

\section{Results}

\begin{figure*}[!t]
\centering
\mbox{\includegraphics[scale=0.7]{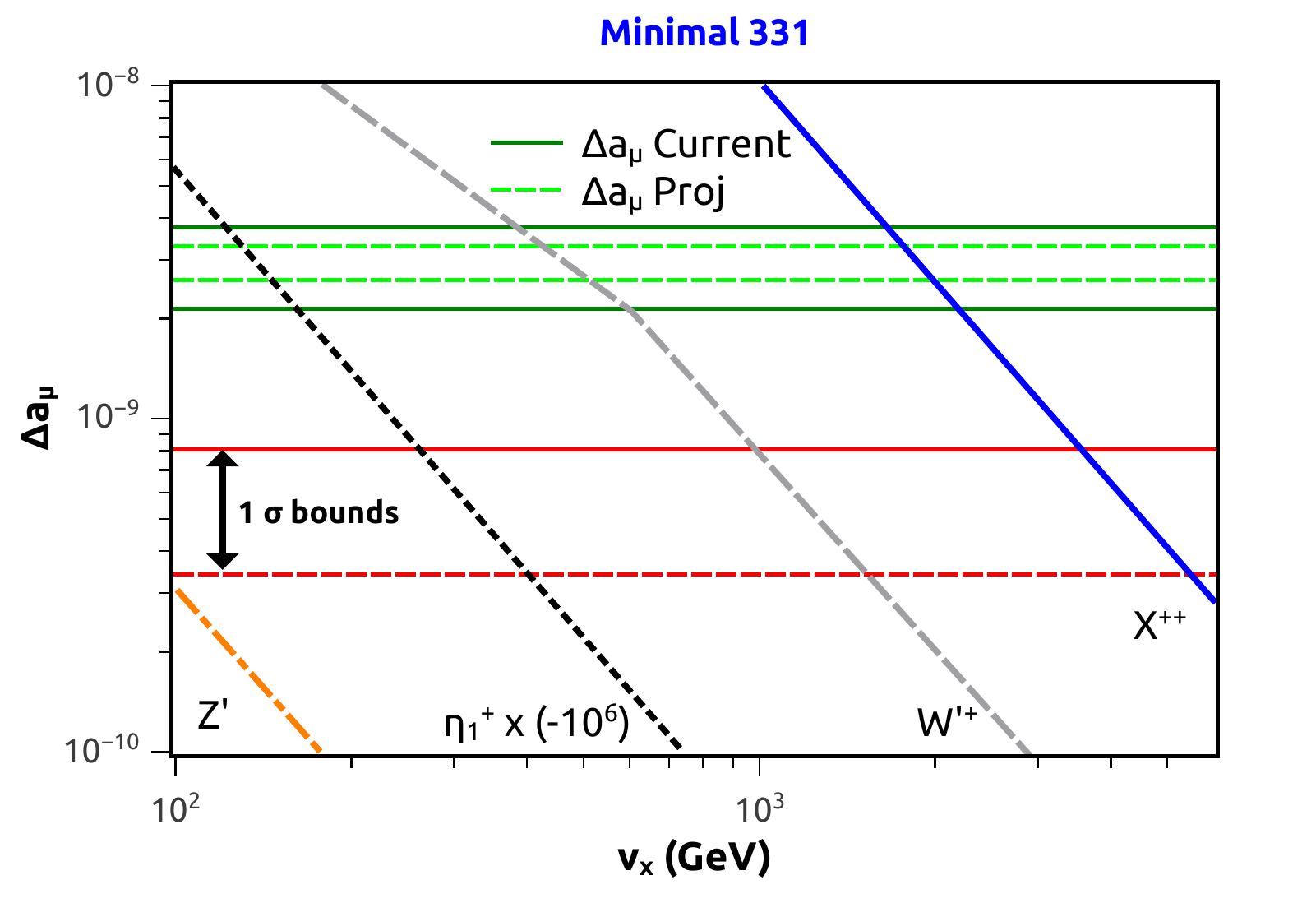}}
\caption{Individual contributions to \gmu. The solid (dashed) green horizontal lines represent the current (projected) sensitive to \gmu. The solid (dashed) horizontal red lines are the current and projected $1\sigma$ bound in case the anomaly is resolved otherwise. After summing up all individual corrections we find that the scale of symmetry breaking that reproduces \gmu\, is quite small and excluded by current data. Moreover, we have derived a current (projected) $4$~TeV ($5.8$~TeV) limit on the scale of symmetry breaking. Since this model is valid only up to 5~TeV, we conclude that the upcoming g-2 experiment at Fermilab will be able to undoubtedly exclude this model.}
\label{Graph1}
\end{figure*}

\begin{figure*}[!h]
\centering
\mbox{\includegraphics[scale=0.7]{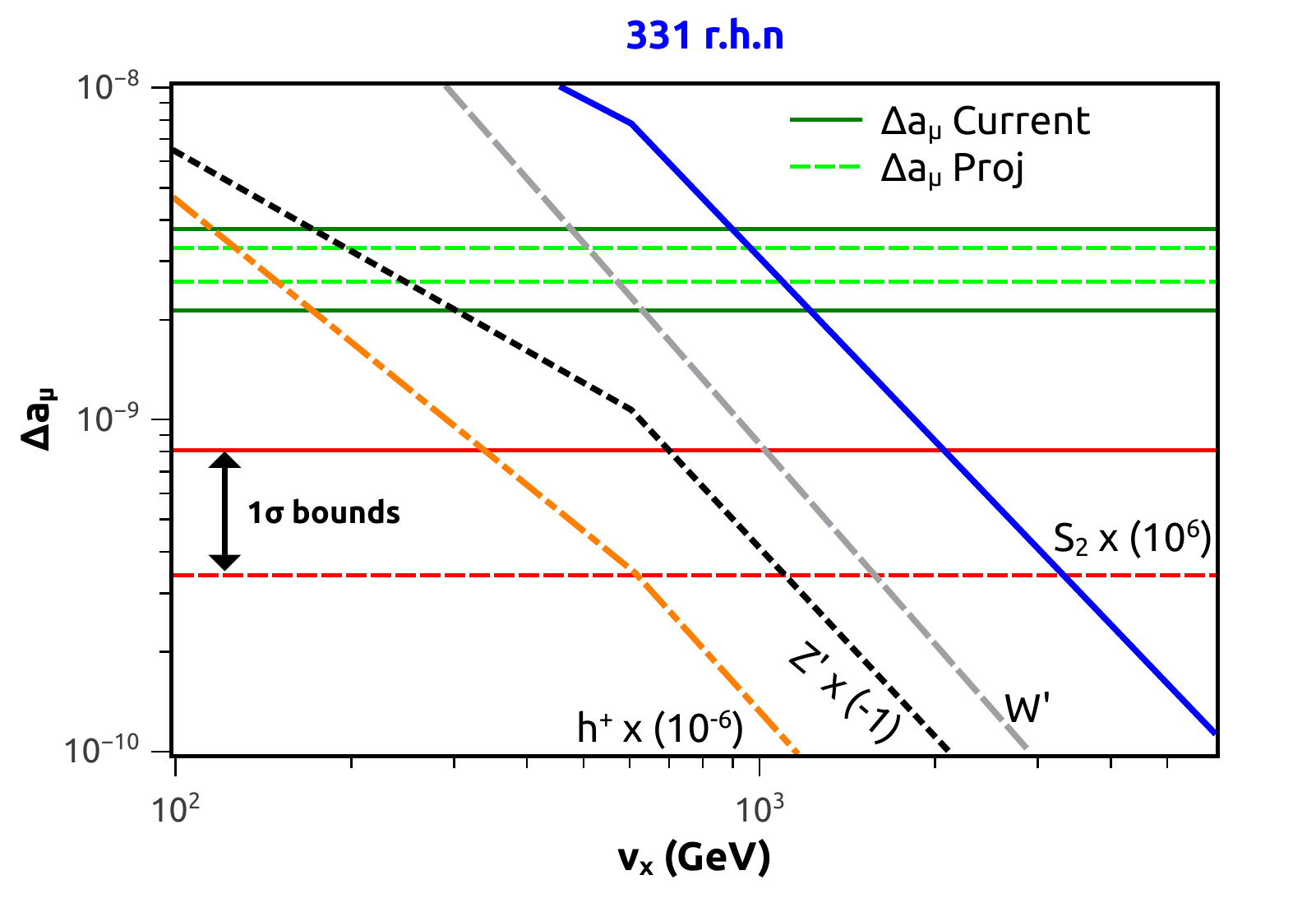}}
\caption{Individual contributions \gmu\, in the 331 r.h.n. Adding up all contributions we conclude that this model cannot explain \gmu\, excess because it requires a scale of symmetry breaking that is already ruled out by current collider,  dark matter experiments and electroweak precision data. Additionally, a current (projected) constraint of $1$~TeV ($1.5$~TeV) might be posed. 
 }
\label{Graph2}
\end{figure*}

\begin{figure*}[!h]
\centering
\mbox{\includegraphics[scale=0.7]{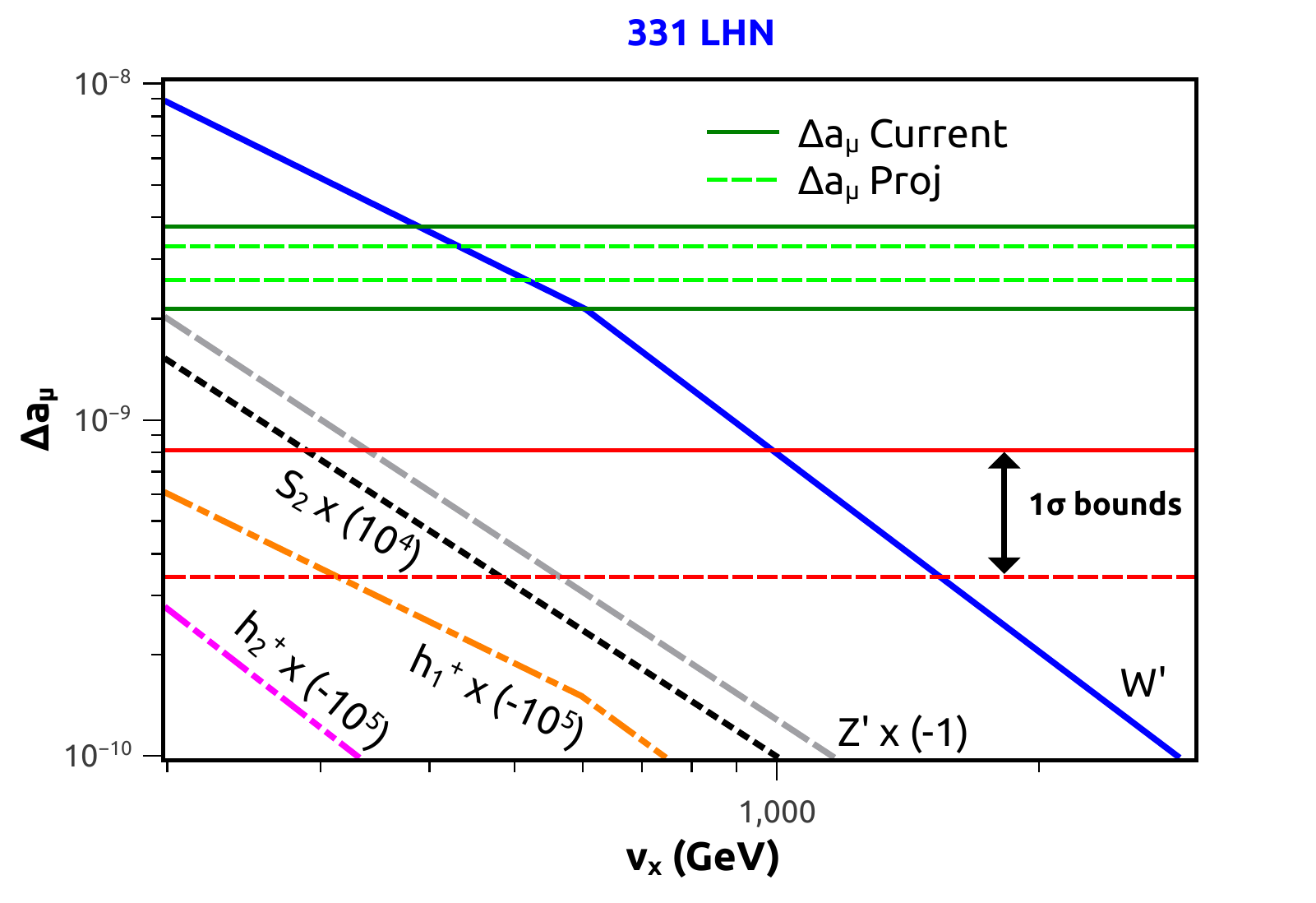}}
\caption{Individual contributions to the muon magnetic moment in the 331 model with heavy leptons as function of the scale of symmetry breaking. With $M_N=1$~GeV, a scale of symmetry breaking of much smaller than $1$~TeV would be needed to accommodate \gmu which is already excluded by dark matter, collider and electroweak bounds. Hence the 331LHN can not provide an explanation for the anomaly.}
\label{Graph3}
\end{figure*}

\begin{figure*}[!h]
\centering
\includegraphics[scale=0.7]{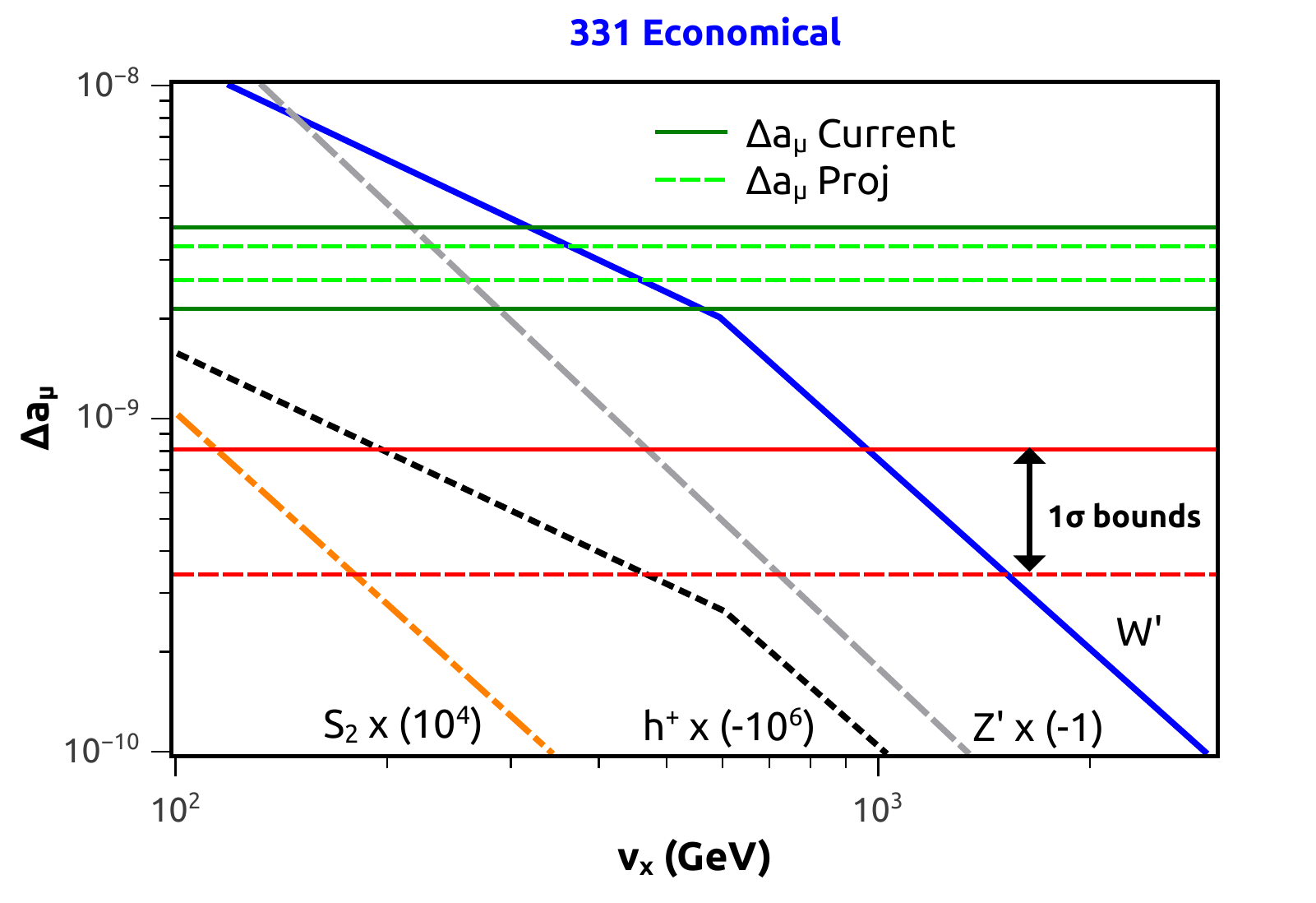}
\caption{Individual contributions from the 331 Economical model. We conclude from the right panel that the overall correction to \gmu\, is negative and small. Thus the model is excluded as an explanation to \gmu. Moreover, a current (projected) limit of $900$~GeV ($1.3$~TeV) can placed on the scale of symmetry breaking.} \label{Graph4}
\end{figure*}

\begin{figure*}[!h]
\centering
\mbox{\includegraphics[scale=0.7]{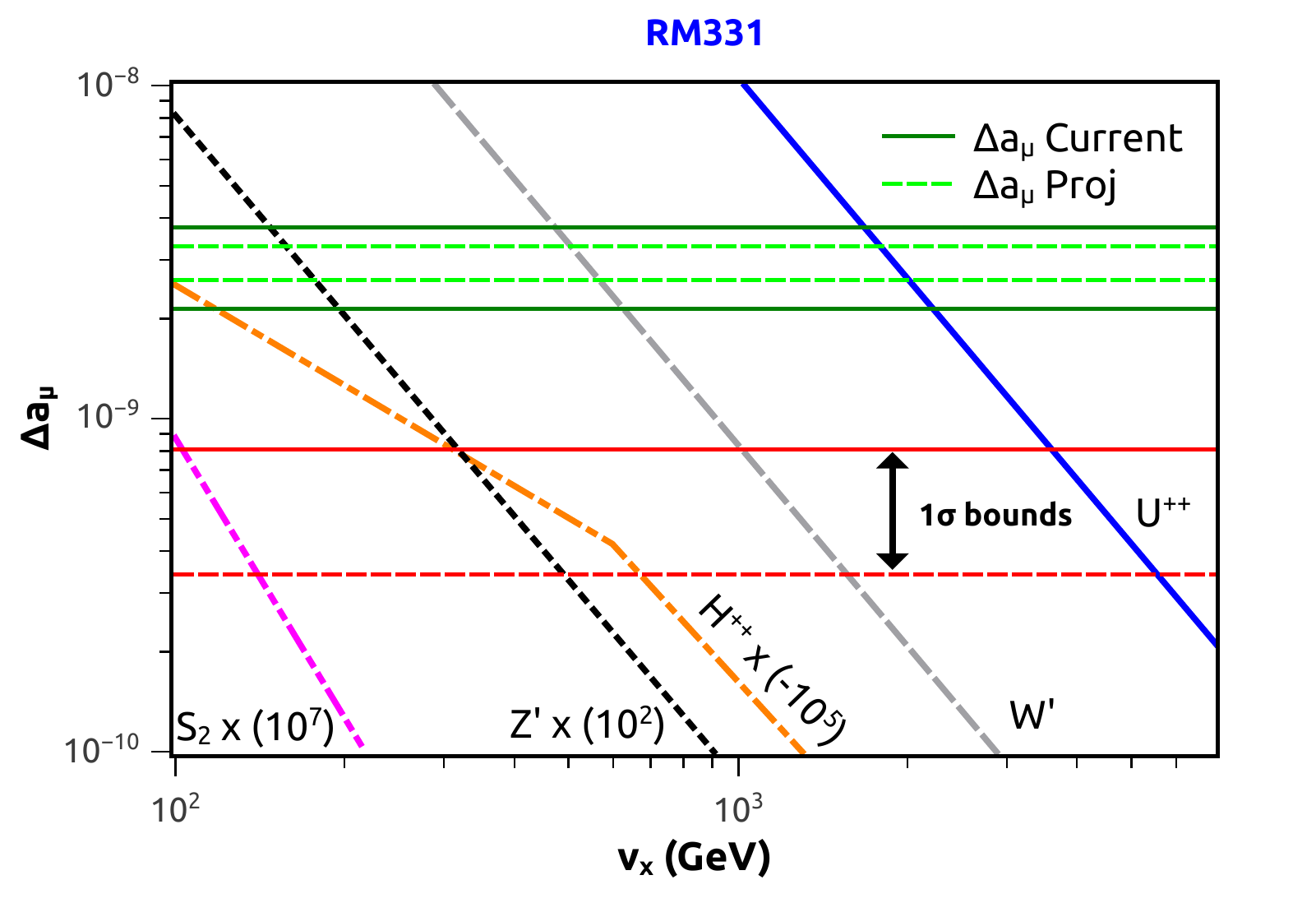}}
\caption{Individual contributions to \gmu in the RM331 as a function of the scale of symmetry breaking. We conclude with $v_{\chi}\sim 2$~TeV the model can accommodate the \gmu\, excess while being consistent with existing bounds. If the anomaly is otherwise resolved a current limit of 4 TeV will be placed, and projected of $5.8$~TeV will be achieved, ruling out the entire model due to the Landau Pole at $5$~TeV.}
\label{Graph5}
\end{figure*}

In this section we present our results taking into account all corrections to the muon magnetic moment stemming from all 331 models previously discussed keeping mind the current constraints. We emphasize that master integrals for computing all these contributions are given in the Appendix. In general contributions coming from scalar particles are suppressed, contributions from neutral gauge bosons can be sizeable with positive or negative sign depending on the relative magnitude of the vector and vector-axial couplings, and doubly charged gauge bosons corrections are large. We presented our numerical results in Figs.\ref{Graph1}-\ref{Graph6}. There the solid (dashed) horizontal green lines delimit the current and projected sensitive of \gmu\, experiments and the region of parameter space which a given model accommodate the reported muon magnetic moment. Below those, the solid (dashed) red lines represent the current (projected) $1\sigma$ bounds that might be placed on the models in case the muon magnetic moment is otherwise resolved. Hereunder we discuss the results shown in Figs.\ref{Graph1}-\ref{Graph6} for the six 331 models. Our limits are summarized in Table \ref{table2}.

\subsection{331 Minimal} 

In Fig.\ref{Graph1} for $v_{\eta}=174$~GeV, we show the numerical results for
the individual contributions to \gmu\. Different values of $v_{\eta}$ produce the same
conclusions since the value of the vev only alters the contribution from the scalars which is negligible. We have not included the neutral scalar contribution because it is also suppressed by the muon mass squared. We display the corrections in
terms of the scale of symmetry breaking because the particles masses have different
dependencies with the scale of symmetry breaking. Therefore, if one plots the results as a function of the masses, the conclusions would be misleading, since the contributions in terms of the particle masses are not on equal footing.

The solid (dashed) green horizontal lines represent the current (projected) sensitivities to \gmu. The solid (dashed) horizontal
red lines are the current and projected $1\sigma$ bound in the case the anomaly is not resolved by any means but this model. The charged scalar correction, which is negative, has been multiplied by $(-10^6)$ to be shown in the graph.
We conclude that the doubly and singly charged vector bosons contributions are the most relevant ones and that for $ v_{\chi}=2$~TeV the model in principle could account for the excess. Albeit, the current LHC bounds rule out $v_{\chi}<3.6$~TeV, and thus this model can be decisively excluded as an explanation of the \gmu\, anomaly. 

Moreover, stringent bounds can be derived  after summing up all contributions. We find that if the anomaly persists, a current (projected) $1\sigma$ limit of 4~TeV (5.8~TeV) can be placed on the scale of symmetry breaking of the model. Since this model is valid only up to 5~TeV or so, we conclude that the upcoming g-2 experiment at Fermilab
will be able to undoubtedly exclude this model. We emphasize that
this conclusion is irrefutable. No fine-tuning or different
parameter choices can remedy this because the main contributions
come from gauge bosons, whose interactions are determined by the gauge group.

\subsection{331 r.h.n}

In Fig.\ref{Graph2} we exhibit the individual contributions to \gmu\, coming from the 
331 r.h.n model as a function of the scale of symmetry breaking. We see that the singly charged (W') and neutral ($Z^{\prime}$) gauge bosons corrections
are the leading ones. We have multiplied the neutral scalar and charged scalar contributions by ($10^6$) and $-1$ respectively to depict them in the graph. Differently from the minimal 331 model, the $Z'$ now gives a negative correction to \gmu\, due to the magnitude of the vector and axial couplings as explained previously.
The $Z'$ contribution has been multiplied by minus one to show it in the plot. We can already notice from the plot that a rather small scale of symmetry breaking is needed to explain the \gmu\, and combining all individual corrections we an overall current (projected) limit of $1$~TeV (1.5)~TeV in the scale of symmetry breaking. Although, current collider, dark matter experiments and electroweak precision data firmly exclude scale smaller than $7.5$~TeV. Hence, the current and projected $1\sigma$ lower bounds on this model are well below the existing ones, this model is excluded as a potential candidate to explain \gmu result in light of the current limits.

\subsection{331 LHN}

In Fig.\ref{Graph3} we exhibit  the individual contributions as a function of the scale of symmetry breaking. Equivalently to the 331 r.h.n the singly charged and neutral gauge bosons are the most relevant corrections to \gmu. The scalars contributions have been multiplied by $10^6$ factor so we could depict them in the panel. The $Z^{\prime}$ contribution is negative and whereas the $W^{\prime}$ is positive but it strongly depends on the mass of the neutral fermion. We have adopted $M_N >1$~GeV. The diagram in question is shown in Fig.\ref{feymann10}. In general the  contribution stemming from $W^{\prime}$ is quite different when its mass is close to the neutral fermion mass. Since we interested in the regime which the scale of symmetry breaking is large than $1$~TeV, i.e $M_W^{\prime} > 330$~GeV, the results from any value of $M_N \ll 330$~GeV are similar to the 331 r.h.n model, where a low scale of symmetry breaking, less than 1TeV, is need to accommodate \gmu\, and the current (projected) limit found is 1 TeV (1.5 TeV) on the scale of symmetry breaking. However, such scale is severely excluded by dark matter, collider and electroweak bounds. In case the neutral fermion mass lies in the TeV scale the overall correction to \gmu\, is dwindled since we are suppressing the leading one ($W^{\prime}$ contribution). In summary, the model similarly to the 331 r.h.n, cannot accommodate the muon anomalous magnetic moment, while obeying the current limits.

\subsection{331 Economical}

In Fig.\ref{Graph4} we depict the individual contributions to \gmu along with the current and projected sensitivities as in previous plots. We have multiplied some of the individual contributions by constants to show them in the plot, namely: neutral scalar $S_2 \times (10^6)$, charged scalar $h_1^+  \times (-10^6)$ and $Z^{\prime} \times (-1)$. One can straightforwardly conclude that the $W^{\prime}$ and $Z^{\prime}$ corrections are the leading ones. It is clear from the figure that the scale of symmetry breaking ($\sim 800$~GeV) required to reproduce the measured \gmu is fiercely ruled out by LHC limits that prohibits scales smaller than $7.5$~TeV. Anyway, we find a current (projected) limit of $1$~TeV in the scale of symmetry breaking of the model. In conclusion, the Economical 331 model cannot
accommodate \gmu. 

\subsection{RM331}

In Fig.\ref{Graph5} we display the individual contributions to \gmu in the RM331 model. We conclude that doubly  charged and singly charged vector bosons are the leading ones.
We we see that a scale of symmetry breaking of $\sim 2$~TeV can explain
the \gmu\, excess. Such energy scale is consistent with the aforementioned limits, since the FCNC ones are sensitive to the parametrization scheme used in the hadronic sector. After summing up all individual corrections we find that in case the anomaly is otherwise resolved,  a current lower bound of 4 TeV, and projected
lower bound of $6$~TeV can be placed on the scale of symmetry breaking of the model. Notice that this model is within current sensitivity of the next generation of \gmu\, experiments. Since this model, similar to the 331 minimal model, is valid up to 5 TeV only due to the Landau Pole, the RM331 might be excluded in the foreseeable future.

\subsection{331 Exotic Leptons}

\begin{figure*}[!t]
\centering
\mbox{\includegraphics[scale=0.7]{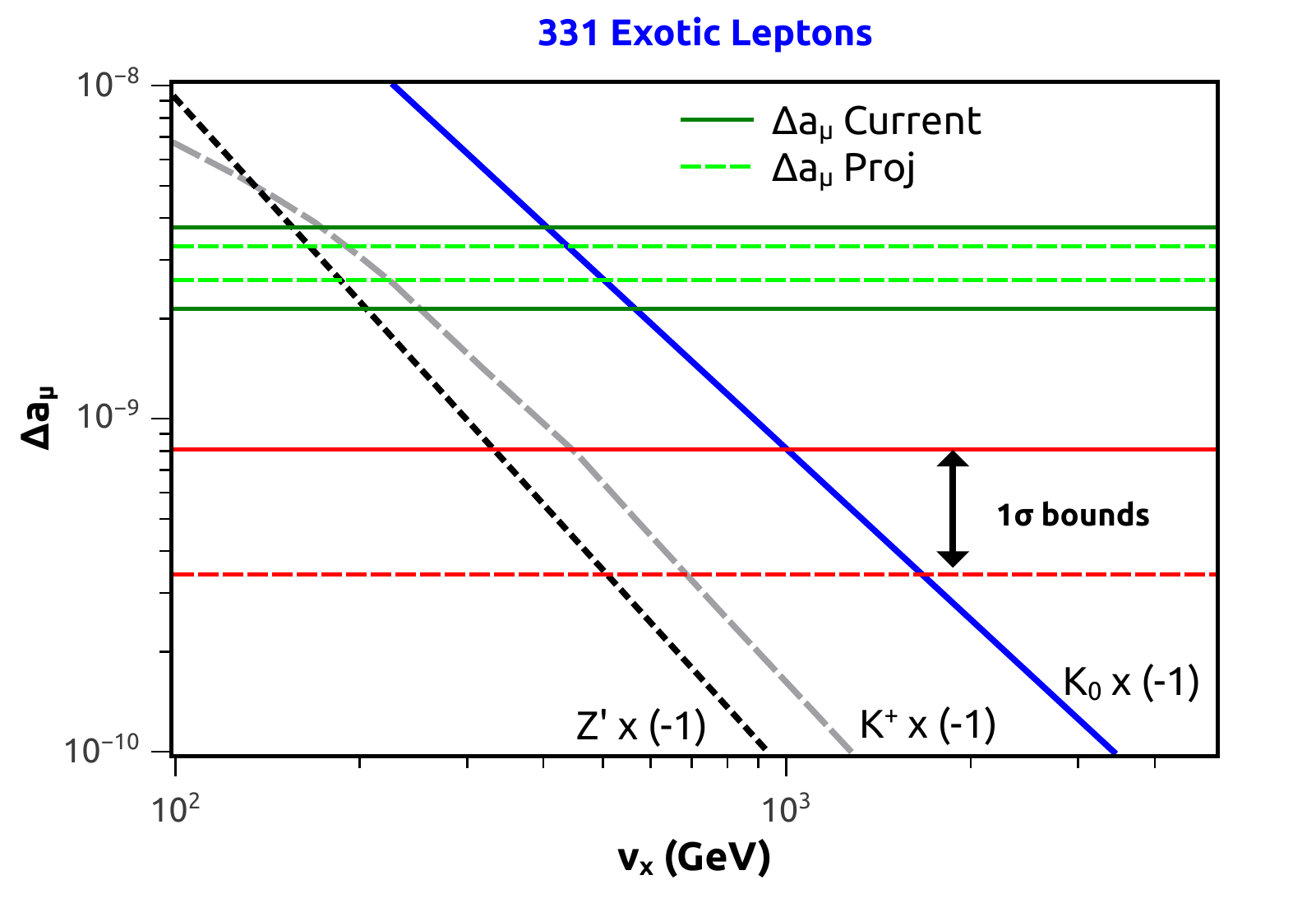}}
\caption{Individual contributions to \gmu in the 331 model with exotic
 leptons. Summing up all corrections we find that overall contribution is negative and small. We could still derive a current (projected) bound of $1$~TeV ($1.8$~TeV) in case the \gmu anomaly is otherwise resolved.}
\label{Graph6}
\end{figure*}

In the Fig.\ref{Graph6} we exhibit the individual corrections from the neutral and charged gauge bosons according to Eq.(\ref{exotcontri}) for exotic leptons (charged leptons) masses of $M_E =1$~TeV. As we discussed previously, the contributions from the scalars have been ignored since they are negligible. One can easily conclude that the $K^0$ correction, which is negative, is the most relevant. Since all corrections are negative the model cannot accommodate the muon magnetic moment excess. Adding up all corrections, we find that the overall contribution is small and negative. However, we can still draw a bound, assuming the anomaly has been otherwise explained, because the overall contribution would still have to lie within the error bars. Hence, we find a current bound of 1~TeV and a projected one of 1.8~TeV in case the \gmu anomaly is otherwise resolved. We emphasize that our limits are not very sensitive to the masses of the exotic leptons. 

\begin{table}[t]
{\begin{tabular}{|c|c|}				
\hline
Model & \gmu\, Limit \\
\hline  
331 Minimal & Current: $V_{\chi} \geq 4$~TeV\\
        &  Projected: $V_{\chi} \geq 5.8$~TeV \\
\hline
331 r.h.n & Current: $V_{\chi} \geq 1$~TeV\\
        &  Projected: $V_{\chi} \geq 1.5$~TeV \\
\hline         
331 LHN & Current: $V_{\chi} \geq 1$~TeV\\
        &  Projected: $V_{\chi} \geq 1.5$~TeV \\
\hline         
331 Economical & Current: $V_{\chi} \geq 900$~GeV\\
        &  Projected: $V_{\chi} \geq 1.3$~TeV \\   
\hline         
RM331 & Current: $V_{\chi} \geq 4$~TeV\\
        &  Projected: $V_{\chi} \geq 5.8$~TeV \\
\hline         
331 Exotic Leptons & Current: $V_{\chi} \geq 1$~TeV\\
        &  Projected: $V_{\chi} \geq 1.8$~TeV \\ 
\hline                                     
\end{tabular}
}
\caption{Limits on the scale of symmetry breaking ($V_{\chi}$) of the 331 models using current and projected sensitivity \gmu\, experiments.}
\label{table2}
\end{table} 	



\section{Conclusions}

We have computed contributions to the muon magnetic moment
stemming from the main 331 models in the literature and derived bounds summarized in Table \ref{table2}. We exploited
the complementarity among collider, dark matter and electroweak
constraints to outline which models are able to address
\gmu. Moreover, we derive stringent $1\sigma$ limits on the scale
of symmetry breaking of those models whenever possible.

We concluded that the {\it Minimal 331 model} cannot explain the \gmu\, excess. We find a robust limit on the
scale of symmetry breaking of $4$~TeV, which can be translated into $M_{Z^{\prime}} > 2.4$~TeV.
As far as we know this is the strongest bound in the $Z^{\prime}$ mass in the literature.
Additionally, we show that the upcoming g-2 experiment at Fermilab will be able to undoubtedly exclude this model.

As for the {\it 331 r.h.n}, we find this model incapable of accommodating measured \gmu\, because it requires a scale of symmetry breaking that is already ruled out by collider and dark matter data, and place a current (projected) $1\sigma$ bound of $1$~TeV ($1.5$~TeV) on the scale of the 331 symmetry breaking.

Regarding the {\it 331LHN}, with $M_N=1$~GeV, a scale of symmetry breaking much smaller than $1$~TeV is needed to address the \gmu\, excess, which is already excluded by current collider and direct dark matter detection. Similarly for $M_N>1$~GeV. A current (projected) $1\sigma$ bound of $1$~TeV ($1.5$~TeV) on the scale of the 331 symmetry breaking was found.

Concerning the {\it 331 Economical model}, due to the large
cancelation between the $W^{\prime}$ and $Z^{\prime}$ corrections
the total contribution is small and requires $v_{\chi} < 1$~TeV to
explain the reported \gmu. This low scale of symmetry breaking is again ruled out by current data. A current (projected) $1\sigma$ bound of $900$~GeV ($1.3$~TeV) on the scale of the 331 symmetry breaking was derived.

As for the {\it RM331} model, we observed that a scale of symmetry breaking of $\sim 2$~TeV could explain the \gmu\, excess, while being consistent with other constraints. If instead, the anomaly is otherwise resolved, a current limit of 4~TeV, and projected of $5.8$~TeV can be placed on the scale of symmetry breaking of the model. Since this model has a Landau pole at 5 TeV (similar to the Minimal 331 model), the RM331 might be entirely excluded in the next generation of experiments.

Lastly, {\it the 331 model with exotic leptons}, predicts a small and negative contribution to \gmu\, regardless of how massive the exotic leptons are. Thus it cannot accommodate the reported \gmu\,. A current (projected) bound of $1$~TeV ($1.8$~TeV) on the scale of symmetry breaking of the model was placed.

\section*{Acknowledgement}
The authors thank Alex Dias, Eugenio Del Nobile for useful discussions. HNL is supported  by the Vietnam National Foundation for Science and Technology Development (NAFOSTED) under grant number 103.01-2014.51. R.M thanks COLCIENCIAS for financial support. FQ is partly supported by US Department of Energy Award SC0010107 and the Brazilian National Counsel for Technological and Scientific Development (CNPq).

\section{Appendix}

In this Appendix we present master integrals for computing \gmu\, stemming from all particles discussed in this work.

\subsection*{Neutral Scalar}
\label{neutralscalar}
Neutral scalars in general can have scalar ($g_{s1}$) and pseudo-scalar ($g_{p1}$) couplings which shift $(g-2)_\mu$ through Fig.\ref{feymann6} by
\begin{eqnarray}
&&
\Delta a_{\mu} (\phi) = \frac{1}{8\pi^2}\frac{m_\mu^2}{ M_{\phi}^2 } \int_0^1 dx \frac{g_{s1}^2 \ P_{s1}(x) + g_{p1}^2 \ P_{p1}(x) }{(1-x)(1-\lambda^2 x) +\lambda^2 x}
\label{scalarmuon1}
\end{eqnarray}where $\lambda=m_{\mu}/M_{\phi}$ and,
\begin{eqnarray}
P_{s1}(x) & = &  x^2 (2 -x),\nonumber\\
P_{p1}(x) & = &  - x^3,
\label{scalarmuon2}
\end{eqnarray}which gives us,
\begin{eqnarray}
\Delta a_{\mu} (\phi)& & = \frac{1}{4\pi^2}\frac{m_\mu^2}{ M_{\phi}^2 }\left[  g_{s1}^2\left(  \ln \left(\frac{ M_{\phi} }{m_{\mu}} \right) -\frac{7}{12}\right) \right. \nonumber\\
+ && \left. g_{p1}^2\left( - \ln \left(\frac{ M_{\phi} }{m_{\mu}} \right) +\frac{11}{12}\right) \right]
\label{scalarfinal}
\end{eqnarray}
The result in Eq.(\ref{scalarfinal}) is for general neutral scalars with scalar and pseudo-scalar couplings in the regime $M_{\phi} \gg m_{\mu}$. Note that neutral scalars are also bounded by LEP searches for four-lepton contact interactions. For $M_{\phi} > \sqrt{s}$ these bounds require $g/M_{\phi} <2.5 \times 10^{-4}{\rm  GeV^{-1}}$ \cite{Queiroz:2014zfa}.

\subsection*{Singly Charged Scalar}
\label{singlychargedS}

A general Lagrangian involving singly charged scalars with scalar ($g_{s1}$) and pseudo-scalar ($g_{p1}$) couplings which gives rise to the $g-2_{\mu}$ correction according to Fig.\ref{feymann1},
\begin{eqnarray}
&&
\Delta a_{\mu} (H^+) = \frac{1}{8\pi^2}\frac{m_\mu^2}{ M_{H^+}^2 } \int_0^1 dx \frac{g_{s2}^2 \ P_{s2}(x) + g_{p2}^2 \ P_{p2} (x) }{\epsilon^2 \lambda^2 (1-x)(1-\epsilon^{-2} x) + x}\nonumber\\
\label{scalarmuon4}
\end{eqnarray}where 
\begin{eqnarray}
P_{s2}(x) & = &  -x (1-x)(x+\epsilon) \nonumber\\
P_{p2}(x) & = & -x (1-x)(x-\epsilon)
\label{scalarmuon5}
\end{eqnarray}with $\epsilon = m_{\nu}/m_{\mu}$ and $\lambda= m_{\mu}/M_{H^+}$, which results in,
\begin{eqnarray}
\Delta a_{\mu} (H^+) &&= \frac{1}{4\pi^2}\frac{m_\mu^2}{ M_{H^+}^2 }\left[  g_{s2}^2\left(-\frac{ m_{\nu} }{4 m_{\mu}} -\frac{1}{12}\right)\right. \nonumber\\
+ && \left. g_{p2}^2 \left( \frac{ m_{\nu} }{4m_{\mu}} -\frac{1}{12}\right) \right]
\label{scalarmuon6}
\end{eqnarray}
Eq.(\ref{scalarmuon6}) holds even if there was a charge conjugation matrix (C) as in same charged scalar contributions in the minimal 331 model presented in Eq.(\ref{chargedscal}).

\subsection*{Doubly Charged Scalar}

A doubly charged scalar  contributes to \gmu through the diagrams Fig.\ref{feymann9}-\ref{feymann8}.  From each diagram we find, respectively,
\begin{eqnarray}
&&\Delta a_{\mu} (H^{\pm \pm})  = \nonumber\\
&& \frac{-q_H}{2 \pi^2} \left( \frac{m_\mu}{M_{H^{\pm \pm}}}\right)^2 \int^1_0 dx \frac{g_{s4}^2 P_s (x) + g_{p4}^2 P_p(x)}{ \lambda^2 x^2 + (1-2\lambda^2)x + \lambda^2 }+\nonumber\\
&  & \frac{-q_f}{2 \pi^2} \left( \frac{ m_\mu}{M_{H^{\pm \pm}}}\right)^2 \int^1_0 dx \frac{g_{s4}^2 P^{\prime}_s (x) + g_{p4}^2 P^{\prime}_p(x) }{ \lambda^2 x^2 + (1-x)}
\label{scalarmuon7}
\end{eqnarray}where
\begin{eqnarray}
P_{s4} (x) & = x^3-x\,\,\,\,; P^{\prime}_s &=  2x^2- x^3 \nonumber\\
P_{p4} (x) & = x^3 -2x^2 +x\,\,\,\,; P^{\prime}_p & =  - x^3 
\label{scalarmuon8}
\end{eqnarray}and $\lambda = m_{\mu}/M_{H^{++}}$, $q_H=-2$ is the electric charge of the doubly charged scalar running in the loop, and $q_f=1$ is the electric charge of the muon in the loop. The factor of four in Eq.(\ref{scalarmuon7}) is a symmetry factor due to the presence of two identical fields in the interaction term. This expression simplifies to,
\begin{equation}
\Delta a_{\mu}(H^{++})= \frac{-2}{3} \frac{g_{s4}^2 m_{\mu}^2 }{\pi^2 M_{\phi^{\pm \pm}}^2 } 
\label{scalarmuon9}
\end{equation}when $g_{p4}=\pm g_{s4}$ and $M_{\phi^{\pm \pm}} \gg m_{\mu}$.
In the setup where either of the above conditions fail the integral in Eq.(\ref{scalarmuon7}) is most easily solved numerically. 

\subsection*{Charged Lepton}

Charged leptons corrects \gmu through Fig.\ref{feymann7}. The contribution is given by,

\begin{eqnarray}
&&
\Delta a_{\mu} (E) = \frac{1}{8\pi^2}\frac{m_\mu^2}{ M_{K^0}^2 } \int_0^1 dx \frac{g_{v6}^2 \ P_{v6}(x) + g_{a6}^2 \ P_{a6} (x) }{(1-x)(1-\lambda^{2} x) +\epsilon^2 \lambda^2 x}\nonumber\\
\label{leptonmuon4}
\end{eqnarray}where 
\begin{eqnarray}
P_{v6}(x) & = &  2x(1-x)(x-2(1-\epsilon))+\lambda^2(1-\epsilon)^2x^2(1+\epsilon-x) \nonumber\\
P_{a6}(x) & = &  2x^2(1+x+2\epsilon)+\lambda^2(1+\epsilon)^2x(1-x)(x-\epsilon)
\label{leptonmuon5}
\end{eqnarray}with $\epsilon = M_{E}/m_{\mu}$ and $\lambda= m_{\mu}/M_{K^0}$.
Therefore the contribution of a generic, singly charged lepton mediated by a neutral vector is found to be
\begin{eqnarray}
&&
\Delta a_{\mu} (E) = \frac{1}{4\pi^2}\frac{m_\mu^2}{ M_{K^0}^2 }\left\lbrace  g_{v6}^2\left[\frac{ M_{E} }{m_{\mu}} -\frac{2}{3}\right] + g_{a6}^2 \left[ -\frac{ M_E}{m_{\mu}} -\frac{2}{3}\right] \right\rbrace, \nonumber\\
\label{leptonmuon6}
\end{eqnarray}in the  $M_K^0 \gg M_{E}$ limit.  Outside of this limit, one should solve Eq.(\ref{leptonmuon4}) numerically, using the public Mathematica code in \cite{Queiroz:2014zfa}, for instance.

\subsection*{Neutral Vector}
Contribution from a new neutral gauge boson, such as a $Z'$, in shown in Fig.\ref{feymann5} and is given by,
\begin{eqnarray}
&&
\Delta a_{\mu} (Z^{\prime}) = \frac{m_\mu^2}{8\pi^2 M_Z^{\prime 2} }\int_0^1 dx \frac{g^2_{v9} P_{v9}(x)+ g^2_{a9} P_{a9}(x) }{(1-x)(1-\lambda^2 x) +\lambda^2 x},\nonumber\\
\label{vectormuon1}
\end{eqnarray} where $\lambda = m_{\mu}/M_{Z^{\prime}}$ and
\begin{eqnarray}
P_{v9}(x) & = & 2 x^2 (1-x) \nonumber\\
P_{a9}(x) & = & 2 x(1-x)\cdot (x-4)- 4\lambda^2 \cdot x^3.
\label{vectormuon2}
\end{eqnarray}
These integrals simplify to give a contribution of 
\begin{equation}
\Delta a_{\mu}(Z^{\prime}) = \frac{m_{\mu}^2}{4 \pi^2 M_Z^{\prime 2}}\left(\frac{1}{3}g^2_{v9} - \frac{5}{3}g^2_{a9}\right)
\label{vectormuon3}
\end{equation}in the limit $M_{Z^{\prime}} \gg m_{\mu}$.
This is the contribution of the $Z^{\prime}$ to the muon anomalous magnetic moment. 
In the regime $M_Z^{\prime} > \sqrt{s}$ LEP has placed a 95\% C.L upper bound of $g_{v9}/M_Z^{\prime} < 2.2\times 10^{-4}{\rm GeV^{-1}}$ for $g_{v9}=g_{a9}$. This limit excludes the possibility of a single $Z^{\prime}$ boson to be the solution to the \gmu\, anomaly \cite{Queiroz:2014zfa}.

\subsection*{Singly Charged Vector} 

Their contributions to \gmu\,  are depicted in Fig.\ref{feymann2} and Fig.\ref{feymann10} reads,
\begin{eqnarray}
&&
\Delta a_{\mu} (W^{\prime}) = \frac{1}{8\pi^2}\frac{m_\mu^2}{ M_{V^+}^2 } \int_0^1 dx \frac{g_{v10}^2 \ P_{v10}(x) + g_{a10}^2 \ P_{a10} (x) }{\epsilon^2 \lambda^2 (1-x)(1-\epsilon^{-2} x) + x},\nonumber\\
\label{vectormuon4}
\end{eqnarray}where 
\begin{eqnarray}
P_{v10}(x) & = &  2x^2(1+x-2\epsilon)+\lambda^2(1-\epsilon)^2 x(1-x)(x+\epsilon) \nonumber\\
P_{a10}(x) & = &  2x^2(1+x+2\epsilon)+\lambda^2(1+\epsilon)^2 x(1-x)(x-\epsilon),\nonumber\\
\label{vectormuon5}
\end{eqnarray}with $\epsilon = m_{\nu}/m_{\mu}$ and $\lambda= m_{\mu}/M_{W^{\prime}}$. This simplifies to
\begin{eqnarray}
&&
\Delta a_{\mu} (W^{\prime}) = \frac{1}{4\pi^2}\frac{m_\mu^2}{ M_{W^{\prime}}^2 } \left[g_{v10}^2 \left( \frac{5}{6} - \frac{m_{\nu}}{m_{\mu}}\right)  + g_{a10}^2 \left(  \frac{5}{6} + \frac{m_{\nu}}{m_{\mu}} \right) \right],\nonumber\\
\label{vectormuon6}
\end{eqnarray}in the regime $M_{W^{\prime}} \gg m_{\mu}$. 

\subsection*{Doubly Charged Vector} 

The doubly-charged vector boson contribution to \gmu\,is exhibited in Figs.\ref{feymann3}-\ref{feymann4}, and are given by,
\begin{eqnarray}
&& \Delta a_{\mu} (U^{\pm \pm}) = \nonumber\\
&& \frac{1}{\pi^2}\left( \frac{m_\mu}{M_{U^{\pm \pm}}}\right)^2\int_0^1 dx \frac{g_{v11}^2 P_{v11}(x) + g_{a11}^2 P_{a11} (x) }{\lambda^2(1-x)^2 + x }\nonumber\\
&  &  \frac{-1}{2\pi^2}\left( \frac{m_\mu}{M_{U^{\pm \pm}}}\right)^2\int_0^1 dx \frac{g^2_{v12} P_{v12}^{\prime}(x)+ g^2_{a12} P_{a12}^{\prime} (x) }{(1-x)(1-\lambda^2 x) + \lambda^2 x},\nonumber\\
\label{Vcontri}
\end{eqnarray}where $\lambda = m_{\mu}/M_{U^{\pm \pm}}$, and
\begin{eqnarray}
P_{v11}(x) & = & 2 x^2(x-1) \nonumber\\
P_{a11}(x) & = & 2 x^2(x+3)+4 \lambda^2 \cdot x (1-x)(x-1), \nonumber\\
P_{v12}^{\prime}(x) & = & 2 x (1-x)\cdot x \nonumber\\
P_{a12}^{\prime}(x) & = & 2 x(1-x)\cdot (x-4)- 4\lambda^2 \cdot x^3.
\end{eqnarray}
Hence the total doubly-charged vector contribution is given by,
\begin{eqnarray}
\Delta a_{\mu} (U^{\pm \pm})&=& \frac{ m_{\mu}^2}{\pi^2 M_{U^{\pm \pm}}^2}\left( \frac{-2}{3}g_{v12}^2 + \frac{16}{3}g_{a12}^2 \right)
\label{doublyvector}
\end{eqnarray} 

\end{document}